\RequirePackage[2020-02-02]{latexrelease}
\documentclass[aps,prb]{revtex4}
\usepackage{amsmath,amssymb}
\usepackage{graphicx}
\usepackage{dcolumn}
\usepackage{bm}
\usepackage{color}
\usepackage{relsize}
\newcommand{\mbb}{\mathbb}
\newcommand{\mc}{\mathcal}

\newcommand{\tet}{\texttt}

\begin{document}
\title{Polarizability and plasmons in pseudospin-1 gapped materials with a flat band}
\author{
Liubov Zhemchuzhna$^{1,2,3}$\footnote{E-mail contact: lzhemchuzhna@mec.cuny.edu, lzhemchuzhna@fordham.edu},
Andrii Iurov$^{1}$\footnote{E-mail contact: aiurov@mec.cuny.edu, theorist.physics@gmail.com},
Godfrey Gumbs$^{3,4}$, 
and
Danhong Huang$^{5}$}

\affiliation{
$^{1}$Department of Physics and Computer Science, Medgar Evers College of City University of New York, Brooklyn, NY 11225, USA\\ 
$^{2}$Department of Physics \& Engineering Physics, Fordham University, Bronx, NY 10458, USA\\
$^{3}$Department of Physics and Astronomy, Hunter College of the City University of New York, 695 Park Avenue, New York, New York 10065, USA\\ 
$^{4}$Donostia International Physics Center (DIPC), P de Manuel Lardizabal, 4, 20018 San Sebastian, Basque Country, Spain\\ 
$^{5}$Space Vehicles Directorate, US Air Force Research Laboratory, Kirtland Air Force Base, New Mexico 87117, USA
}

\date{\today}

\begin{abstract}
The collective electronic properties of various types of pseudospin-$1$ Dirac-cone materials with a flat band and finite bangaps in their energy spectra are the subject of our reported investigation.  Specifically,  we have calculated the dynamical polarization, plasmon dispersions as well as their decay rates due to Landau damping. Additionally, we present closed-form analytical expressions for the wave function overlaps for both the gapped dice lattice and the Lieb lattice. The gapped dice lattice is a special case of the more general $\alpha$-${\cal T}_3$ model since its band structure is symmetric and the flat band remains dispersionless.  On the other hand, the Lieb lattice has a flat band which appears at the  lowest point of its conduction band. Our results for these two cases exhibit unique features in the plasmon spectra and their damping regions, which have never been reported in previous studies. For example, the particle-hole modes of a Lieb lattice appear as finite-size regions,  while the plasmon modes exist only in a region with small wave numbers but an extended range of frequencies.  
\end{abstract}

\maketitle

\section{Introduction} 
\label{sec1}

Recent research on almost all aspects of electronic property in novel two-dimensional materials has become a crucial subject as well as one of the most actively pursued directions  in condensed-matter physics, chemistry, photonics and quantum electronics. This is a consequence of unique electron dynamics in graphene whose properties have been considered to be among  the most significant discoveries in this century.\,\cite{geim2009graphene,geim2007rise,meyer2007structure} Shortly after discovery of plain graphene, researchers made another effort to fabricate and investigate other relevant candidates, such as graphene with a finite bandgap between its valence and conduction bands,   novel materials with anisotropic and/or tilted Dirac cones as well as indirect band gaps,\,\cite{wild2022optical,milicevic2019type,trescher2015quantum}  new materials with Rashba spin-orbit coupling,\,\cite{sun2023determining, shitrit2013spin, wang2005plasmon,beyer2023rashba} massive anisotropic electronic states,\,\cite{mojarro2021optical,zhang2024quantum} twisted bilayers,\,\cite{navarro20233,naumis2021reduction,hou2024strain,hu2023josephson} semi-Dirac materials,\,\cite{xiong2023optical,islam2018driven,carbotte2019optical,mondal2022topology} and {\em etc\/}. The band structures in several types of these materials also demonstrate valley and spin-polarized electronic properties\,\cite{tabert2013valley} and very unusual low-energy electronic spectra as well.   Remarkably, anisotropic dispersion, accompanied by an energy gap in the band structure, of Dirac-cone materials could also be induced and controlled by applying an external off-resonance irradiation.\,\cite{iurov2024floquet,dey2019floquet,dey2018photoinduced,iurov2022floquet,kristinsson2016control,kibis2010metal,ibarra2019dynamical}
\medskip 
\par

Among such novel Dirac materials, those, involving a flat or dispersionless band in energy spectrum, are particularly interesting and stand out because of their highly unusual electronic,\,\cite{iurov2020klein,islam2023properties,iurov2019peculiar,tamang2023probing,tamang2023orbital,anwar2020interplay,
islam2023effect,gorbar2019electron,illes2017klein} collective,\,\cite{malcolm2016frequency,oriekhov2020rkky} optical, magnetic and transport\,\cite{iurov2020quantum,oriekho2023quantum} properties. Importantly, such materials are all connected to the so-called $\alpha$-${\cal T}_3$ model described by a hexagon with an additional hub atom in its center, Lieb lattice,\,\cite{slot2017experimental,mukherjee2015observation,jo2012ultracold} Kagome lattice\,\cite{guo2009topological,lee2024atomically,mojarro2023topological,wang2024dispersion} and others, respectively.   
\medskip 
\par

Plasmon, or quantum quasi-particles representing collective charge-density fluctuations within a conducting solid, is one of the important properties of low-dimensional materials.  The motivation for this work is  to gain further insight into    the dynamical polarization, plasmon dispersions as well as their decay rates.  With the use of angle-resolved photoemission spectroscopy (ARPES) or high resolution electron energy loss spectroscopy (HREELS),  A quantitative study of plasmon modes, including dispersion and dephasing (lifetime or damping), has been performed for graphene with zero and finite bandgap,\,\cite{wunsch2006dynamical,hwang2007dielectric,polini2008plasmons,pyatkovskiy2008dynamical,politano2014plasmon,yan2012infrared} silicene and germanene,\,\cite{tabert2014dynamical} hetero-structures,\,\cite{gumbs2015nonlocal} hybrid systems,\,\cite{iurov2017effects,mojarro2022hyperbolic} multi-layers, fullerenes,\,\cite{solov2005plasmon, gumbs2014strongly, ju1993excitation} carbon nanotubes\,\cite{bondarev2009strong, bondarev2009optical} and nanoribbons, as well as other materials,\,\cite{dey2022dynamical,wang2005plasmon,shitrit2013spin,brongersma2015plasmon,sarma1981collective,mikhailov2011theory,backes1992dispersion,agarwal2014long} at either zero or finite temperatures\,\cite{sarma2013intrinsic,iurov2022finite}. Particularly, studies of plasmon modes have been conducted for flat-band Dirac materials\,\cite{malcolm2016frequency,kajiwara2016observation}.
In all these studies, a special attention has been drawn towards plasmon damping and stability\,\cite{koseki2016giant, petrov2017amplified,gumbs2015tunable,simon1983inhomogeneous} since only a weakly-damped plasmon mode can represent a quasi-particle.\,\cite{woessner2015highly} Meanwhile, magneto-plasmons and energy-dispersion of electrons under a magnetic field have also been explored fully.\,\cite{balassis2020magnetoplasmons,nimyi2022landau,roldan2009collective,roldan2011theory,gumbs2014revealing} Moreover, studies on dynamics of plasmon modes have been performed for tilted and anisotropic lattices, such as 1T'-MoS$_2$ and $8$-{\em pmmn} borophene\,\cite{yan2022anomalous, balassis2022polarizability, torbatian2021hyperbolic, sadhukhan2017anisotropic, jalali2018tilt,hayn2021plasmons}, graphene, silicene and $\alpha$-${\cal T}_3$ based nanoribbons\,\cite{gomez2016plasmon, karimi2017plasmons, fei2015edge, iurov2021tailoring, brey2007elementary,tan2021anisotropic} in addition to related collective and transport properties, such as optical and Boltzmann conductivities.\,\cite{wareham2023optical, iurov2023optical, stauber2013optical, yan2023highly,iurov2018temperature, tan2021anisotropic, gomes2021tilted} 
\medskip 
\par
The remainder of this paper is organized as follows. In Secs.~\ref{sec2} and \ref{sec3}, we calculate the low-energy Hamiltonian, electron energy spectrum and wave functions in both gapped dice and Lieb lattices by means of electronic properties in well-known gapped pseudospin-$1$ Dirac materials with a flat band. Meanwhile, we discuss similarity and difference in electron dynamics of these two likewise materials. The consequent Sec.~\ref{sec4} is devoted to defining and calculating dynamical polarization functions, plasmon dispersions and dampings in these materials. Specifically, we discuss the zero-frequency limit of the polarization function and static screening, which is known to be important in calculations of the Boltzmann conductivity. Furthermore, we show how the density of states could be employed for  reliably estimating  the plasmon frequency in the long wavelength limit for gapped dice and Lieb lattices, as well as other known Dirac materials. In conjunction with the theory, numerical results are presented and discussed in detail for displaying photo-excited electron dynamics in these two different materials. Finally, the concluding remarks and summary statements are given in Sec.~\ref{sec5}. 
\medskip

\section{Finite-gap dice lattice: electronic states and band structure}
\label{sec2}

\begin{figure} 
\centering
\includegraphics[width=0.5\textwidth]{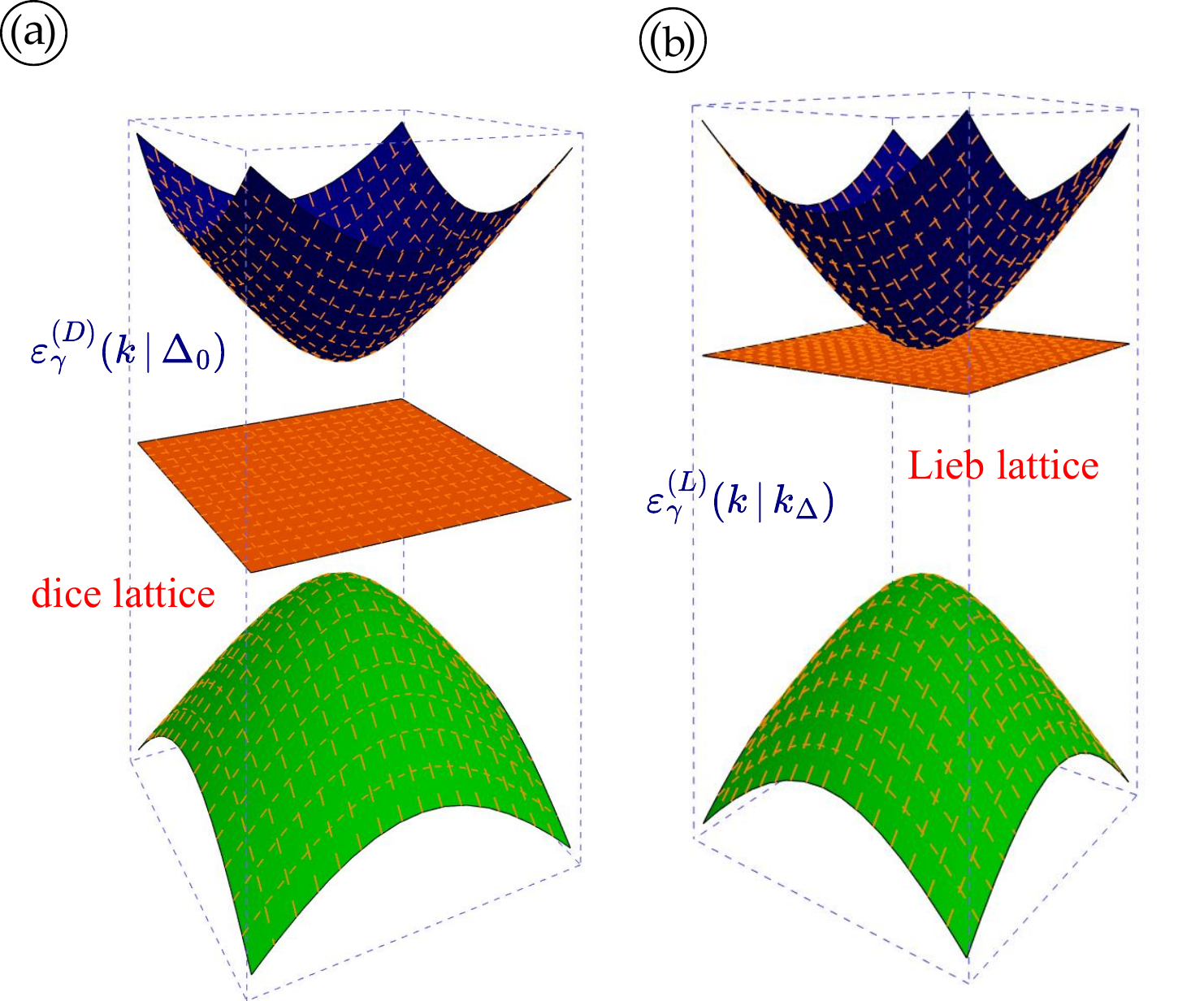}
\caption{(Color online) Schematics of the low-energy dispersions (specturm) for a dice lattice and a Lieb lattice. In both cases, there exist three energy bands, and one of the bands is flat (dispersionless). However, the location of the flat band is essentially different for the two lattices.}
\label{FIG:1}
\end{figure}

Mathematically speaking, a dice lattice corresponds to the $\phi \rightarrow \pi/4$ limit of an $\alpha$-${\cal T}_3$ model, which was addressed and investigated extensively in Refs.\,[\onlinecite{oriekhov2022optical,weekes2021generalized}]. Specifically, the Hamiltonian for a gapped dice lattice is supplemented by an extra bandgap-related term, {\em i.e.\/}, 

\begin{equation}
\label{S1z06}
\hat{\Sigma}^{(1)}_{z} = \left[
\begin{array}{ccc}
 1 & 0 & 0 \\
 0 & 0 & 0 \\
 0 & 0 & -1
\end{array}
\right] \ , 
\end{equation}
while the Hamiltonian itself is given by   
 
\begin{equation}
\label{nhd26D}
\hat{\mc{H}}_{d} (\mbox{\boldmath$k$} \, \vert \, \tau) = \frac{\hbar v_F}{\sqrt{2}} \, \left[
\begin{array}{c c c}
\sqrt{2} \, \Delta_0 & k^\tau_{-} & 0 \\
k^\tau_{+} & 0 & k^\tau_{-} \\
0 & k^\tau_{+} & -\sqrt{2} \, \Delta_0
\end{array}
\right] \ .
\end{equation}
The energy dispersions for the gapped dice and Lieb lattices are presented in Fig.~\ref{FIG:1}.

\medskip 
Here, in contrast to general $\alpha$-${\cal T}_3$ model, the eigenvalue equation, corresponding to Eq.\,\eqref{nhd26D}, gives rise to a symmetric band structure with two equal bandgaps and an unaffected flat band right in the middle between a valence and conduction band. Therefore, by including the extra term in Eq.\,\eqref{S1z06}, the flat band will be deformed and receives a finite $k$-dispersion for all $\alpha$-${\cal T}_3$ materials with a finite gap, expect for a dice lattice having $\phi = \pi/4$.  
\medskip 

The energy dispersions, corresponding to Hamiltonian \eqref{nhd26D}, are given by $\varepsilon^{\tau}_{\gamma=0} (\mbox{\boldmath$k$}\,\vert\,\Delta_0)  \equiv 0$, and 

\begin{equation}
\varepsilon^{\tau}_{\gamma=\pm 1} (\mbox{\boldmath$k$}\,\vert\,\Delta_0) = \gamma \sqrt{k^2 + \Delta_0^2} \ , 
\end{equation}
which is valley $(\tau)$ independent. The corresponding wave functions for $\gamma=\pm 1$ of this gaped dice lattice are calculated as 

\begin{equation}
{\bf \Psi}_{\gamma = \pm 1}^{\tau} (\mbox{\boldmath$k$}\,\vert\,\Delta_0) = \frac{1}{2} \, \left\{
\begin{array}{c}
\left[ 1 + \frac{\gamma \Delta_0}{E_{\Delta_0}(k)} \right] \, \tet{e}^{- i \Theta_{\bf k}} \\[0.05cm]
\sqrt{2} \gamma \, \sqrt{1 - \left( \frac{\Delta_0}{E_{\Delta_0}(k)} \right)^2} \\[0.05cm]
\left[  1 - \frac{\gamma \Delta_0}{E_{\Delta_0}(k)}  \right] \, \tet{e}^{i \Theta_{\bf k}}
\end{array}
\right\} \ ,
\end{equation}
which is also $\tau$ independent, where $E_{\Delta_0}(k) =  \sqrt{k^2 + \Delta_0^2} = \varepsilon^\tau_{\gamma=1} (\mbox{\boldmath$k$}\,\vert\,\Delta_0)$. For the flat band ($\gamma=0$), however, its $\tau$-independent wave function is obtained as 

\begin{equation}
{\bf \Psi}_{\gamma = 0}^{\tau} (\mbox{\boldmath$k$}\,\vert\,\Delta_0) = \frac{1}{\sqrt{2}} \, \sqrt{1 - \left( \frac{\Delta_0}{E_{\Delta_0}(k)} \right)^2} \, \left\{
\begin{array}{c}
 - \tet{e}^{- i \Theta_{\bf k}} \\[0.05cm]
\sqrt{2} \frac{\Delta_0}{\sqrt{E_{\Delta_0}(k)^2 - \Delta_0^2 }} \\ [0.05cm]
 \tet{e}^{i \Theta_{\bf k}}
\end{array}
\right\} \, .
\end{equation}
In addition, the overlap integral $\mathfrak{O}_{\gamma, \gamma'}^{\tau} (\mbox{\boldmath$k$},\mbox{\boldmath$q$}\,\vert\,\Delta_0)$, required for calculating the polarization function, is defined by

\begin{equation}
\label{OI-1}
\mathfrak{O}_{\gamma, \gamma'}^{\tau}(\mbox{\boldmath$k$},\mbox{\boldmath$q$}\,\vert\,\Delta_0) = \Big | \, \Big \langle  {\bf \Psi}_{\gamma }^{\tau} (\mbox{\boldmath$k$}\,\vert\,\Delta_0) \, \Big \vert \,
{\bf \Psi}_{\gamma'}^{\tau} (\mbox{\boldmath$k$}+\mbox{\boldmath$q$}\,\vert\,\Delta_0)
 \Big \rangle \, \Big |^2\ .
\end{equation}
\medskip

\medskip

Explicitly, by using Eq.\,\eqref{OI-1}, the overlaps for Dirac-cone states with $\gamma = \pm 1$ are calculated as 

\begin{eqnarray}
\nonumber
&& \mathfrak{O}_{\gamma, \gamma'}^{\tau} (\mbox{\boldmath$k$},\mbox{\boldmath$q$}\,\vert\,\Delta_0) = \frac{1}{4} \, \left\{ 
1 + \gamma \gamma' \, \cos \left( \beta^\delta_{\,{\bf k},{ \bf q}} \right) \, \sqrt{
\left[1 - \left( \frac{\Delta_0}{E_{\Delta_0}(k)} \right)^2 \right] \, \left[1 - \left( \frac{\Delta_0}{E_{\Delta_0}(\mbox{\boldmath$k$}+\mbox{\boldmath$q$})} \right)^2 \right]
}
\right\}^2 \\
\nonumber 
& + & \frac{\Delta_0^2}{2 \, E_{\Delta_0}(k) \, E_{\Delta_0}(\mbox{\boldmath$k$}+\mbox{\boldmath$q$})} \, \left\{
 \gamma \gamma' + \cos^2 \left( \beta^\delta_{\,{\bf k},{ \bf q}} \right) \,
 \sqrt{
\left[1 - \left( \frac{\Delta_0}{E_{\Delta_0}(k)} \right)^2 \right] \, \left[1 - \left( \frac{\Delta_0}{E_{\Delta_0}(\mbox{\boldmath$k$}+\mbox{\boldmath$q$})} \right)^2 \right]
}
\right\}
\\
\nonumber
& = & \frac{1}{4} \, \left\{ 
1 + \gamma \gamma' \, \cos \left( \beta^\delta_{\,{\bf k},{ \bf q}} \right) \, \frac{k}{E_{\Delta_0}(k)} \, \frac{\vert\mbox{\boldmath$k$}+\mbox{\boldmath$q$}\vert}{E_{\Delta_0}(\mbox{\boldmath$k$}+\mbox{\boldmath$q$})}
\right\}^2 + \frac{\Delta_0^2}{2 \, E_{\Delta_0}(k) \, E_{\Delta_0}(\mbox{\boldmath$k$}+\mbox{\boldmath$q$})} \, \left\{
 \gamma \gamma' + \cos^2 \left( \beta^\delta_{\,{\bf k},{ \bf q}} \right) \, \right. \\
\label{mainOg1}
& \times &
\left. 
\frac{k}{E_{\Delta_0}(k)} \, \frac{\vert\mbox{\boldmath$k$}+\mbox{\boldmath$q$}\vert}{E_{\Delta_0}(\mbox{\boldmath$k$}+\mbox{\boldmath$q$})}
\right\} \, , 
\end{eqnarray}
where $\beta^\delta_{\,{\bf k},{ \bf q}}$ is the angle between two vectors $\mbox{\boldmath$k$}$ and $\mbox{\boldmath$k$}+\mbox{\boldmath$q$}$, so that $\cos \left( \beta^\delta_{\,{\bf k},{ \bf q}} \right) = (k^2 + \mbox{\boldmath$k$}\cdot\mbox{\boldmath$q$})/( k  \, \vert\mbox{\boldmath$k$}+\mbox{\boldmath$q$}\vert)$. Specifically, for $\gamma'=0$ and $\gamma=\pm1$, one finds from Eq.\,\eqref{mainOg1} that 

\begin{eqnarray}
\label{mainOg2}
&& \mathfrak{O}_{\gamma = \pm 1, \gamma' = 0}^{\tau} (\mbox{\boldmath$k$},\mbox{\boldmath$q$}\,\vert\,\Delta_0) = \frac{1}{2} \, \frac{\vert\mbox{\boldmath$k$}+\mbox{\boldmath$q$}\vert}
{E_{\Delta_0}(\mbox{\boldmath$k$}+\mbox{\boldmath$q$})} \, \left\{
\sin^2 \left( \beta^\delta_{\,{\bf k},{ \bf q}} \right)  - \left[ \frac{\Delta_0}{E_{\Delta_0}(k)}  \right]^2 \, \frac{\mbox{\boldmath$k$}\cdot\mbox{\boldmath$q$}}{k \, \vert\mbox{\boldmath$k$}+\mbox{\boldmath$q$}\vert}
\right\} \\
\nonumber 
& = &  \frac{1}{2} \, \frac{\vert\mbox{\boldmath$k$}+\mbox{\boldmath$q$}\vert}{E_{\Delta_0}(\mbox{\boldmath$k$}+\mbox{\boldmath$q$})} \, \left\{
\frac{k^2}{\vert\mbox{\boldmath$k$}+\mbox{\boldmath$q$}\vert^2} - \left( \frac{\mbox{\boldmath$k$}\cdot\mbox{\boldmath$q$}}{ q \, \vert\mbox{\boldmath$k$}+\mbox{\boldmath$q$}\vert}
\right)^2
 - \left[ \frac{\Delta_0}{E_{\Delta_0}(k)} \right]^2  \, \frac{\mbox{\boldmath$k$}\cdot\mbox{\boldmath$q$}}{k \, \vert\mbox{\boldmath$k$}+\mbox{\boldmath$q$}\vert}
\right\} \ ,
\end{eqnarray}
which involves the electron transition from/towards the flat band. 
\medskip

\section{Energy spectrum and electronic states in a Lieb lattice}
\label{sec3}

Physically, besides a finite-gap dice lattice discussed in Sec.\,\ref{sec2}, flat bands can also be realized in some other types of lattices, mostly, optical Lieb and Kagome lattices. A Lieb lattice was observed in a number of existing systems and experimental setups, {\em e.g.\/}, organic materials, optical lattices and waveguides.
Mathematically, however, a Lieb lattice can be viewed as the combination of three displaced square sublattices. The Hamiltonian of a Lieb lattice is generally written as\,\cite{oriekhov2022optical}

\begin{equation}
\label{lieb106} 
\mc{H}^{(\text{\,L})}(\mbox{\boldmath$k$}\, \vert \, k_\Delta) = \hbar v_F \, \left[
\begin{array}{c c c}
 k_\Delta & k_x & 0 \\[0.1cm]
 k_x & - k_\Delta  & k_y \\[0.1cm]
 0 & k_y & k_\Delta 
\end{array}
\right] \ , 
\end{equation}
where the following substitution

\begin{equation}
k_{x,y} \rightarrow \frac{\pi}{a_0} + k_{x,y}
\label{subst}
\end{equation}
is required, and $a_0$ stands for the lattice parameter. 
\medskip 

Before taking the substitution in Eq.\,\eqref{subst}, three energy dispersions can be easily found from the Hamiltonian in Eq.\,\eqref{lieb106} as $\varepsilon^{(\text{\,L})}_{\gamma = \pm 1} (\mbox{\boldmath$k$}\, \vert \, k_\Delta)/\hbar v_F = \gamma \sqrt{ k_\Delta^2 + k_x^2 + k_y^2}$ and $\varepsilon^{(\text{\,L})}_{\gamma = 0} (\mbox{\boldmath$k$}\, \vert \, k_\Delta)/\hbar v_F = k_\Delta$, which could be combined into a single expression, yielding

\begin{equation}
\label{lieb116}
\varepsilon^{(\text{L})}_{\gamma}(\mbox{\boldmath$k$}\, \vert \, k_\Delta) = \hbar v_F \, \left[\delta_{\gamma, 0} \, k_\Delta + \gamma \,(1 - \delta_{\gamma, 0})\, \sqrt{k_\Delta^2 + k^2} \,\right] \, , 
\end{equation}
where $\gamma=0,\,\pm 1$, and $\delta_{\gamma, 0}$ is the Kronecker symbol. Consequently, we obtain three energy subbands, and one of them $(\gamma=0)$ is a flat band.
From Eq.\,\eqref{lieb116}, it is obvious that the flat band in a Lieb lattice is located at a finite energy $\hbar v_F k_\Delta$, which is right next to the lowest point of the conduction band, while for the case of a dice lattice with a finite gap, this flat band is located symmetrically with respect to the valence and conduction bands. 
\medskip 

The corresponding wave functions with respect to energy bands in Eq.\,\eqref{lieb116} are given by 

\begin{eqnarray}
\label{liebwg}
{\bf \Psi}^{(\text{L})}_{\gamma = \pm 1} (\mbox{\boldmath$k$}\, \vert \, k_\Delta) & = & 
\frac{1}{ 2 E_k (E_k - \gamma\hbar v_Fk_\Delta)}
\,
\left\{
\begin{array}{c}
k_x \\
- k_\Delta + \gamma E_k \\
k_y
\end{array}
\right\}\ ,
\end{eqnarray}
and 

\begin{equation}
\label{liebw0}
{\bf \Psi} ^{(\text{L})}_{\gamma = 0} (\mbox{\boldmath$k$}\, \vert \, k_\Delta) =   \frac{1}{k} \, \left\{
\begin{array}{c}
- k_y \\
0\\
k_x
\end{array}
\right\} = \left\{
\begin{array}{c}
- \sin \Theta_{\bf k} \\
0\\
\cos \Theta_{\bf k}
\end{array}
\right\}\ ,
\end{equation}
where $E_k\equiv\hbar v_F\,\sqrt{k_\Delta^2+k^2}$ and $\Delta_0=2\hbar v_Fk_\Delta$ is the bandgap. It is straightforward to verify that eigenstates \eqref{liebwg} and \eqref{liebw0} are orthogonal and normalized.
\medskip

Using the obtained wave functions in Eqs.\,\eqref{liebwg} and \eqref{liebw0}, we calculate their overlaps as defined in Eq.\,\eqref{OI-1}. For isotropic dispersions we can always consider wave vector $q$ aligned to a specific directions, such as $x$-axis. By using this fact, we find

\begin{eqnarray}
\label{mainOgL1}
&& \mathfrak{O}_{\gamma =  \pm 1, \gamma' =  \pm 1 } (\mbox{\boldmath$k$}, \mbox{\boldmath$q$}\,\vert\,k_\Delta) = \frac{
\left\{ 
\mbox{\boldmath$k$} \cdot \mbox{\boldmath$q$}  +  \left(
E_k - \gamma k_\Delta
\right)
\left[
E_k + E_{{\bf k} + {\bf q}} + (\gamma - \gamma') k_\Delta
\right] 
\right\}^2
}{2 E_k \, E_{{\bf k} + {\bf q}} \, \left(
E_k - \gamma k_\Delta
\right) \left(
E_{{\bf k} + {\bf q}} - \gamma' k_\Delta
\right) } \ , 
\end{eqnarray}
and 

\begin{eqnarray}
\label{mainOgL2}
&& \mathfrak{O}_{\gamma = \pm 1, \gamma' = 0}(\mbox{\boldmath$k$}, \mbox{\boldmath$q$}\,\vert\,k_\Delta) = \frac{
\left[ (\mbox{\boldmath$k$} \times \mbox{\boldmath$q$})\cdot\hat{\mbox{\boldmath$e$}}_z \right]^2
}{2 E_k \left(
E_k - \gamma k_\Delta \right)
\left|\mbox{\boldmath$k$} + \mbox{\boldmath$q$} \right|^2 } =  
\frac{
k^2q^2 - (\mbox{\boldmath$k$} \cdot \mbox{\boldmath$q$})^2
}{2 E_k \left(
E_k - \gamma k_\Delta \right)
\left|\mbox{\boldmath$k$} + \mbox{\boldmath$q$} \right|^2 }
\ , 
\end{eqnarray}
where $(\mbox{\boldmath$k$} \times \mbox{\boldmath$q$})\cdot\hat{\mbox{\boldmath$e$}}_z$ represents the $z$-component (out-of-plane component) of the cross product of vectors $\mbox{\boldmath$k$}$ and $\mbox{\boldmath$q$}$. 
\medskip

\section{Polarization function and plasmon dispersions}
\label{sec4}

We now turn our attention to the dynamics of plasmon modes in gapped dice and Lieb lattices, which are among the most crucial collective electronic 
properties of these novel Dirac materials. For this, we first look into calculating the plasmon dispersion, which is determined by the following characteristic equation

\begin{equation}
\epsilon(q, \omega\, \vert \, \Delta_0) =  1 - V_C(q) \, \Pi^{(0)}(q, \omega\, \vert \, \Delta_0) = 0\ ,
\label{eps01}
\end{equation}
where $q$ and $\omega$ are the wave number and frequency of photo-excited electrons, respectively, while $\epsilon(q, \omega\, \vert \, \Delta_0)$ stands for a dielectric function defined within the $q$-$\omega$ plane.  In Eq.\,\eqref{eps01}, $V_C(q)=2 \pi \beta_\epsilon/q = e^2 /2\epsilon_0\epsilon_r q$ is an electron-electron Coulomb potential within a two-dimensional plane, and  
$\beta_\epsilon = e^2/4\pi\epsilon_0 \epsilon_r$ represents the (inverse) dielectric constant of a (dielectric) substrate. The solutions  of the secular equation  $\epsilon(q,\omega|\Delta_0)=0$  for the dielectric function defined in  Eq.\,\eqref{eps01} can in general be written as $\omega=\Omega_{\rm pl}(q)$.  This gives rise to a dispersion relation for the plasmon-mode frequency $ \Omega_{\rm pl}(q)$ as a function of  wave number $q$. The dielectric function in Eq.\,\eqref{eps01} is  given in terms of the polarization function for interacting electrons which in the random-phase approximation (RPA) is $\Pi^{RPA}(q,\omega\vert \, \Delta_0) =
\Pi^{(0)}(q,\omega\vert \, \Delta_0)/
\epsilon(q, \omega\, \vert \, \Delta_0)$. The dielectric  function  is usually complex, like the dynamical polarization function $\Pi^{(0)}(q, \omega\, \vert \, \Delta_0)$ defined in Eq.\,\eqref{Pi00} below. Here, the imaginary part of $\epsilon(q, \omega\, \vert \, \Delta_0)$ or $\Pi^{(0)}(q, \omega\, \vert \, \Delta_0)$ determines the damping region of a plasmon mode within the $q$-$\omega$ plane, {\em i.e.\/}, how likely a collective quasi-particle will decay into single-electron excitations. A stable plasmon mode acquires a very long lifetime, which is expected if both real and imaginary parts of the dielectric function in Eq.\,\eqref{eps01} equal to zero. 
\medskip 
\par

In the following, we will  calculate the noninteracting polarization function which was introduced in Eq.\,\eqref{eps01} for a pseudospin-1 Hamiltonian. This is obtained from the equation of motion for the density matrix in the absence of dissipation  and working to lowest order in the external perturbation.  This provides   the induced density fluctuation in terms of the density-density response function   
$\Pi^{(0)}({\bf r},{\bf r}^\prime, \omega \, \vert \, \phi, \mu(T)) $  determined by four wave functions  which depend on  the coordinate variables  ${\bf r}$ and ${\bf r}^\prime$.  Making use of the fact that the polarization depends on the difference ${\bf r}  -{\bf r}^\prime$, we Fourier transform leading to
----

\begin{equation}
\Pi^{(0)}(q, \omega \, \vert \, \phi, \mu(T)) = \frac{g}{4 \pi^2} \, \int d^2\mbox{\boldmath$k$} \sum\limits_{\gamma,\gamma' = 0 \, \pm 1} \, 
\mathfrak{O}_{\gamma, \gamma'} (\mbox{\boldmath$k$},\mbox{\boldmath$q$}\,  \vert\, k_\Delta) \,
\frac{n_F[\epsilon_\gamma(k\,\vert\,\Delta_0),\,\mu] - n_F[\epsilon_{\gamma'}(\vert \mbox{\boldmath$k$}+\mbox{\boldmath$q$}\vert\,\vert\,\Delta_0),\,\mu]}{ \left(\hbar \omega + i 0^+ \right) + \epsilon_\gamma  (k\,\vert\,\Delta_0) - \epsilon_{\gamma'} (\vert \mbox{\boldmath$k$} +\mbox{\boldmath$q$}\vert\,\vert\,\Delta_0)} \ ,
\label{Pi00}
\end{equation}
where $n_F[\epsilon_\gamma (k\,\vert\, \Delta_0), \mu(T)] = \left\{1 + \tet{exp}[(\epsilon_\gamma (k\,\vert\,\Delta_0) - \mu)/(k_B T)] \right\}^{-1}$ is the Fermi-Dirac distribution function in a thermal-equilibrium state of electrons. The wave-function overlaps $\mathfrak{O}_{\gamma, \gamma'} (\mbox{\boldmath$k$}, \mbox{\boldmath$q$}\,  \vert\, k_\Delta)$ have been calculated in Eqs.\,\eqref{mainOg1} and \eqref{mainOg2} for a dice lattice as well as in Eqs.\,\eqref{mainOgL1} and \eqref{mainOgL2} for a Lieb lattice, separately. 
Here, before we take on rigorous numerical computations for the dynamical polarization function in Eq.\,\eqref{Pi00}, dielectric function in Eq.\,\eqref{eps01} and plasmon-mode dispersions for both gapped dice and Lieb lattices, we would like to introduce first an extremely useful and instructive technique in estimating plasmon dispersion in the long-wavelength limit based on the calculated electron density $n_0$ and the density of states $\rho_d(\mc{E})$ as well. 
\medskip

In general, the density $n_0(T)$ of an electron gas at a finite temperature $T$ can be calculated by 

\begin{equation}
\label{tids}
n_0(T) = \int\limits_0^\infty \rho_d(\mc{E})\, f(\mc{E}) \, d \mc{E} \ , 
\end{equation}
where $\mc{D}(\mc{E})$ is the density of states while $f(\mc{E})$ stands for the occupation function of electrons. 
Specifically, at $T=0\,$K, we find that Eq.\,\eqref{tids} simply reduces to  

\begin{equation}
n_0 =  \int\limits_0^{E_F} \rho_d(\mc{E}) \, d \mc{E}  \, , 
\end{equation}
where $E_F$ represents the Fermi energy of electrons in the system. 
\medskip

For two-dimensional gapped dice and Lieb lattice systems, their density of states $\rho_d(\mbb{E})$ of electrons can be written as  

\begin{equation}
\label{intpe}
\rho_d(\mbb{E})  = \frac{g_s g_v}{(2 \pi)^2} \, \int\limits_0^{\infty} \, k \, d k \, \int\limits_0^{2 \pi}  \, d \Theta_{\bf k} \, \delta \Big[ \mbb{E} - \varepsilon_{\gamma} (\mbox{\boldmath$k$} \, \vert \, \Delta_0)  \Big] \ ,
\end{equation}
which could be used for various Dirac materials, where $g_s$ and $g_v$ represent spin and valley degeneracies, respectively, while $\varepsilon_{\gamma} (\mbox{\boldmath$k$} \, \vert \, \Delta_0)$ is the energy dispersion of electrons in the system. For the simplest case with a graphene, we find that $\rho_d(\mbb{E}) = 2 \mbb{\vert E \vert}/[\pi (\hbar v_F)^2]$. For the gapped graphene with its energy dispersions $\varepsilon_{\lambda} (k\,\vert\,\Delta_0) = \lambda \sqrt{(\hbar v_F k)^2 + \Delta_0^2}$, on the other hand, we obtain 

\begin{equation} 
\label{dosgg}
\rho_d(\mbb{E},\Delta_0) = \frac{1}{\pi} \, \sum \limits_{\lambda = \pm 1}  \frac{2 \mbb{E} }{ \lambda (\hbar v_F)^2} \, \Theta \left( \frac{\mbb{E}}{\lambda}  - \Delta_0 \right) \, , 
\end{equation}
which leads to a charge density at $T=0$\,K, given by 

\begin{equation} 
n_0(\Delta_0) = \frac{E_F^2 - \Delta_0^2}{\pi (\hbar v_F)^2} \, \Theta \left( E_F  - \Delta_0 \right) \ . 
\end{equation} 
Finally, Eq.\,\eqref{dosgg} leads us to the well-known expression for the polarization function 

\begin{equation}
\Pi^{(0)}(q,\omega\,\vert \, \Delta_0) \Big\vert_{q \ll k_F} = \frac{E_F}{\pi} \, 
\sqrt{1 - \left( \frac{\Delta_0}{E_F} \right)^2}
\,\left( \frac{q}{\hbar \omega} \right)^2 \, \Theta \left( E_F  - \Delta_0 \right) \ . 
\end{equation}
\medskip 

In the presence of both a bandgap $\Delta_0$ and a flat band at $\mbb{E} = 0$ ({\em i.e.\/} gapped dice lattice), Eq.\,\eqref{intpe} gives rise to the following expression 

\begin{equation}
\label{rhodd}
\rho^{\,({\rm D})}_d(\mbb{E}\,\vert\,\Delta_0) =  \frac{1}{\pi} \, \sum \limits_{\lambda = \pm 1}  \frac{2 \mbb{E} }{ \lambda (\hbar v_F)^2} \, \Theta \left( \frac{\mbb{E}}{\lambda}  - \Delta_0 \right) + \frac{1}{4 \pi} \,  \int\limits_0^{k_{\text{max}} \longrightarrow \infty} d \left(k^2 \right) \,\delta (\mbb{E})\ ,
\end{equation}
while for a Lieb lattice we acquire 

\begin{equation}
\label{rhodl}
\rho^{\,({\rm L})}_d(\mbb{E}\,\vert\,\Delta_0) =  \frac{1}{\pi} \, \sum \limits_{\lambda = \pm 1}  \frac{2 \mbb{E} }{ \lambda (\hbar v_F)^2} \, \Theta \left( \frac{\mbb{E}}{\lambda}  - \Delta_0 \right) + \frac{1}{4 \pi} \,  \int\limits_0^{k_{\rm max} \longrightarrow \infty} d \left(k^2 \right) \,\delta (\mbb{E} - \Delta_0)\ .
\end{equation}
For both cases in Eqs.\,\eqref{rhodd} and \eqref{rhodl}, their corresponding electron densities are the same, which is calculated as 

\begin{equation} 
n(\Delta_0) = \frac{E_F^2 - \Delta_0^2}{\pi (\hbar v_F)^2} \, \Theta \left( E_F  - \Delta_0 \right) + \frac{1}{4 \pi} \,  \int\limits_0^{k_{\rm max} \longrightarrow \infty} d \left(k^2 \right)  \ . 
\end{equation} 
From the density of states in both Eqs.\,\eqref{rhodd} and \eqref{rhodl}, we find that the last term $\displaystyle{\backsim \int\limits_0^{k_{{\rm max}\longrightarrow \infty}} d \left(k^2 \right)}$ for a flat band is divergent and cannot be properly taken into account ``as it is''. We know that the flat band only affects the long-wavelength limit in the next order of $q$ as 

\begin{eqnarray} 
\label{rigpre}
 \hbar \left[\Omega^{(\alpha)}_p(q) \right]^2 &=& \frac{ 8 E_F^2 \, \beta_c \,\, (\hbar v_F q)}{4 E_F + 4 \left[1 + 12 \alpha^2 (1+\alpha^2)^{-2} \, \hbar v_F q \right]} \\
\nonumber
& = & 2 E_F \, \beta_c \, (\hbar v_F q) - 2 \beta_c \, \left[1 + \frac{12\,\alpha^2}{(1+\alpha^2)^2} \right] \, (\hbar v_F q)^2 + . . . \ ,  
\end{eqnarray} 
where $\beta_c=\sin^2(2\phi)$, $\alpha=\tan(\phi)$ is the relative hopping parameter for the $\alpha$-$\mc{T}_3$ model, which equals $0$ for graphene but $1$ for a dice lattice. Therefore, we get  
$\left[ \alpha^{(D)} \right]^2  \, \left\{ 1 + \left[\alpha^{(D)}\right]^2 \right\}^{-2} = 1/4$ for a dice lattice in Eq.\,\eqref{rigpre}. 
\medskip

Based on the calculated density of states, the plasmon frequency is estimated as

\begin{equation}
\label{lowp1}
\Omega_p (q) \Big|_{q \ll k_F} = \left\{\beta_c \, q \, \hbar v_F \sqrt{\pi n_0}  \right\}^{1/2} \ , 
\end{equation}
which yields the same dependence on the material parameters ({\em e.g.\/}, Fermi energy, Fermi velocity and dielectric constant) as in the exact results for the plasmon dispersion for all known Dirac materials. Using this approach, the long-wavelength plasmons frequencies are calculated as

\begin{equation}
\left[\hbar \Omega_p^{({\rm G})} (q) \right]^2 = 2 \beta_c \, E_F\, \hbar v_F q
\end{equation}
in graphene,

\begin{equation}
\label{ggplas}
\left[\hbar \Omega_p^{(\rm {G.G})} (q) \right]^2 = 2 \beta_c \, \sqrt{E_F^2 - \Delta_0^2} \, \Theta \left( E_F  - \Delta_0 \right) \, \hbar v_F q
\end{equation}
in gapped graphens, and

\begin{eqnarray} 
\left[\hbar \Omega_p^{({\rm S})} (q) \right]^2 & = & 2 \beta_c \, \sqrt{E_F^2 - \left(\Delta_<^2+\Delta_>^2 \right)^2} \,\, \Theta \left( E_F  - \Delta_> \right) \, \hbar v_F q \\
\nonumber 
& + & 2 \beta_c \, \sqrt{E_F^2 - \Delta_<^2} \,\, \Theta \left( E_F  - \Delta_< \right) \Theta \left(\Delta_> - E_F \right) \, \hbar v_F q
\end{eqnarray} 
in silicene with two generally inequivalent bandgaps $\Delta_< \leq \Delta_>$.
\medskip

As a particularly interesting and straightforward result, we can easily predict the plasmon dispersions for an Key-Y patterned graphene with two inequivalent gapless Dirac cones or   two different Fermi velocities $v_{F,1} < v_{F,2}$.\,\cite{andrade2019valley,herrera2020electronic} For this case, its density of states $\rho_d(\mbb{E},\Delta_0)$ is calculated as

\begin{equation} 
\rho_d(\mbb{E},\Delta_0) = \frac{1}{\pi} \, \sum \limits_{\lambda = \pm 1}  \frac{\vert \mbb{E} \vert }{\left[ \hbar v_{(F,\lambda)} q \right]^2} \ . 
\end{equation}
This results in its electron density $n(\Delta_0)$, given by 

\begin{equation} 
n(\Delta_0) = \frac{1}{\pi} \, \sum \limits_{\lambda = \pm 1}  \frac{\vert \mbb{E_F} \vert }{\left[ \hbar v_{(F,\lambda)} q \right]^2}\ . 
\end{equation}
At the same time, its plasmon frequency takes the form

\begin{equation}
\left[\hbar \Omega_p^{({\rm G})} (q) \right]^2 = \beta_c \, E_F \, \sum \limits_{\lambda = \pm 1} \left[ \hbar v_{(F,\lambda)} q \right] \  ,
\end{equation}
which appears as an average between the ``fast'' and ``slow'' Dirac cones. Therefore, we conclude that for the flat-band materials with an energy bandgap ({\em i.e.\/}, Dice and Lieb lattice), the flat band does not play a crucial role for the plasmon spectrum in the long-wavelength limit, and it could be obtained in the ``density-of-states'' approximation as equivalent to that of gapped graphene in Eq.\,\eqref{ggplas}. This can be verified from the rigorous result in Eq.\,\eqref{rigpre} (see Ref.\,[\onlinecite{oriekhov2022optical}]) for dice lattice and $\alpha-\mc{T}_3$ model in the absence of a bandgap $\Delta_0$.  
\medskip 

\begin{figure} 
\centering
\includegraphics[width=0.55\textwidth]{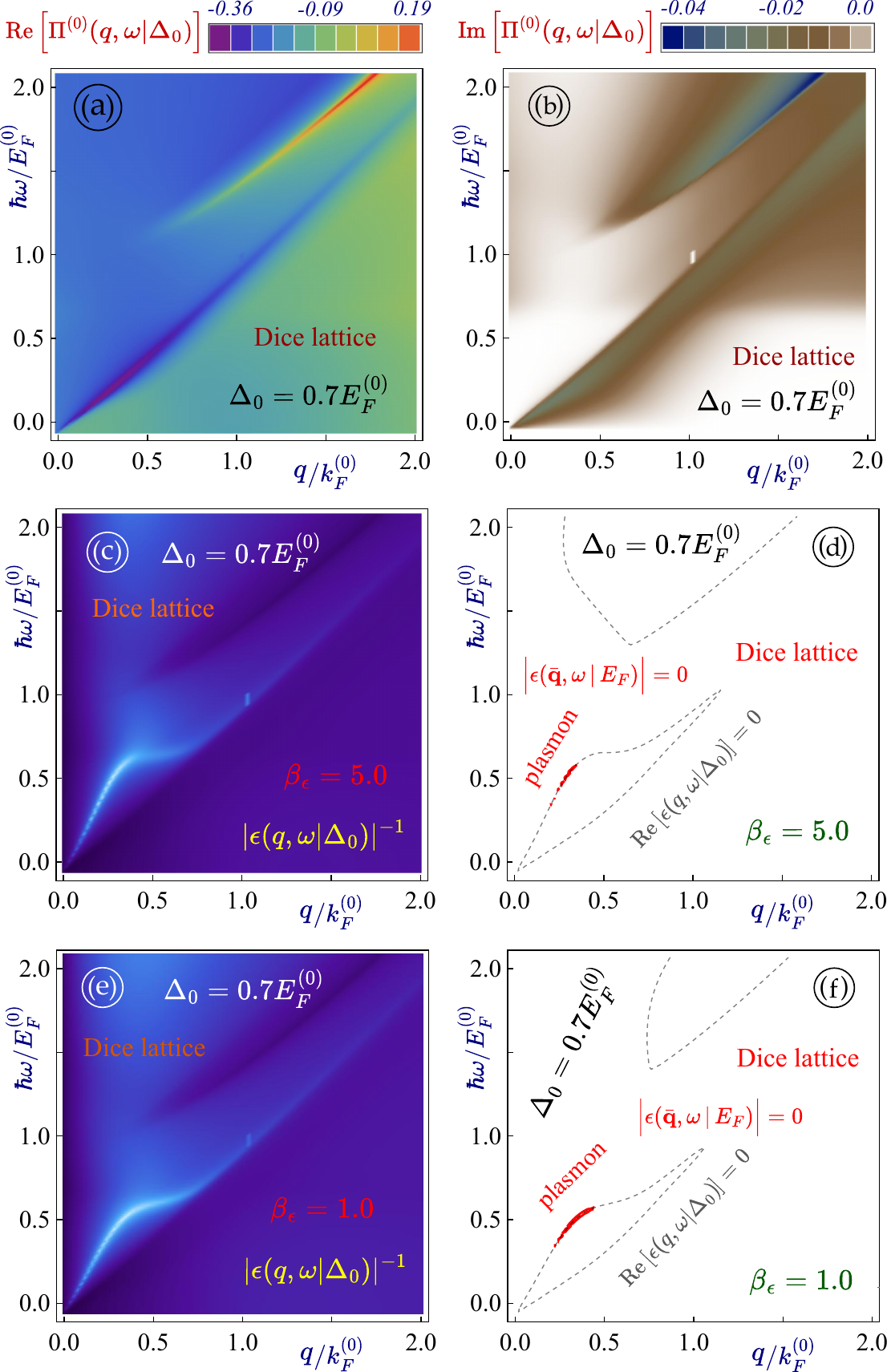}
\caption{(Color online) Polarization function $\Pi^{(0)}(q, \omega\, \vert \, \Delta_0)$ and dielectric function $\epsilon(q,\omega\,\vert\,\Delta_0)$ are shown for a gapped dice lattice with a large badngap $\Delta_0 = 0.7\,E_F^{(0)}$. Top panels $(a)$ and $(b)$ demonstrate the density plots for the real and imaginary parts for $\Pi^{(0)}(q, \omega\, \vert \, \Delta_0)$ as functions of wave vector $q$ and plasmon excitation frequency $\omega$. Plasmon damping and singe-particle excitation regions are shown as areas with a finite imaginary part of $\text{Im}\,[\Pi^{(0)}(q, \omega\, \vert \, \Delta_0)]$ in panel $(b)$. The remaining plots $(c)$-$(f)$ display the plasmon spectrum in a dice lattice, obtained as the zeros of a dielectric function in Eq.\,\eqref{eps01}. The left panels $(c)$ and $(e)$ represent the density plots for inverse dielectric function $\left|1/\epsilon(q, \omega\,\vert \, \Delta_0)\right|$ whose peaks correspond to the plasmon branches, while the right-hand-side plots $(d)$ and $(f)$ present numerically calculated plasmon dispersions, obtained as the zeros of $\left|\epsilon(q, \omega\,\vert \, \Delta_0)\right|$. The middle panels $(c)$ and $(d)$ correspond to inverse dielectric constant $\beta_\epsilon = 1/(4 \pi \epsilon_0 \epsilon_r)  = 5$, while the lower plots $(e)$ and $(f)$ to $\beta_\epsilon = 1$.}
\label{FIG:2}
\end{figure}
\medskip

\begin{figure} 
\centering
\includegraphics[width=0.55\textwidth]{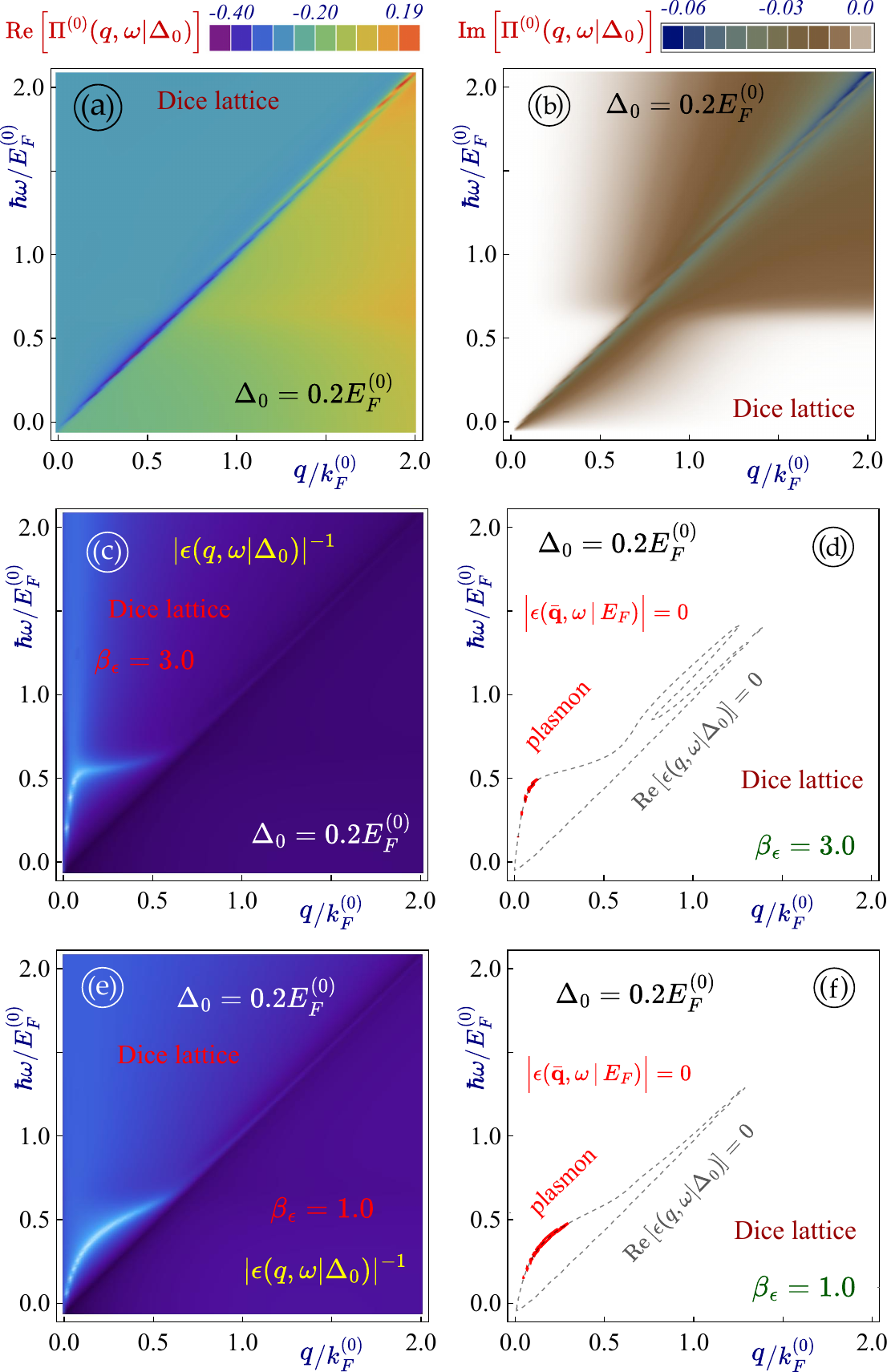}
\caption{(Color online) Polarization function and plasmon spectrum for a gapped dice lattice with a small badngap $\Delta_0 = 0.2\,E_F^{(0)}$. 
Top panels $(a)$ and $(b)$ demonstrate the density plots of the real and imaginary parts of the polarization function $\Pi^{(0)}(q, \omega\,\vert \, \Delta_0)$ as functions of wave vector $q$ and plasmon excitation frequency $\omega$. Plasmon damping and singe-particle excitation regions are shown as areas with a finite imaginary part of $\text{Im}\,[\Pi^{(0)}(q, \omega\,\vert \, \Delta_0)]$ in panel $(b)$. The remaining plots $(c)$-$(f)$ display the plasmon spectrum in a gapped dice lattice, obtained as the zeros of dielectric function in Eq.\,\eqref{eps01}. The left panels $(c)$ and $(e)$ represent the density plots for inverse dielectric function $\left|1/\epsilon(q, \omega\,\vert \, \Delta_0)\right|$ whose peaks correspond to the plasmon branches, while the right-hand-side plots $(d)$ and $(f)$ exhibit numerically calculated plasmon dispersions, obtained as the zeros of  $\vert \epsilon(q, \omega\,\vert \, \Delta_0)\vert$. The middle panels $(c)$ and $(d)$ correspond to inverse dielectric constant $\beta_\epsilon = 1/(4 \pi \epsilon_0 \epsilon_r)  = 5$, while the lower plots $(c)$ and $(d)$ to $\beta_\epsilon = 1$.}
\label{FIG:3}
\end{figure}
\medskip

\section{Numerical Results and Discussion}
\label{added}

\begin{figure} 
\centering
\includegraphics[width=0.55\textwidth]{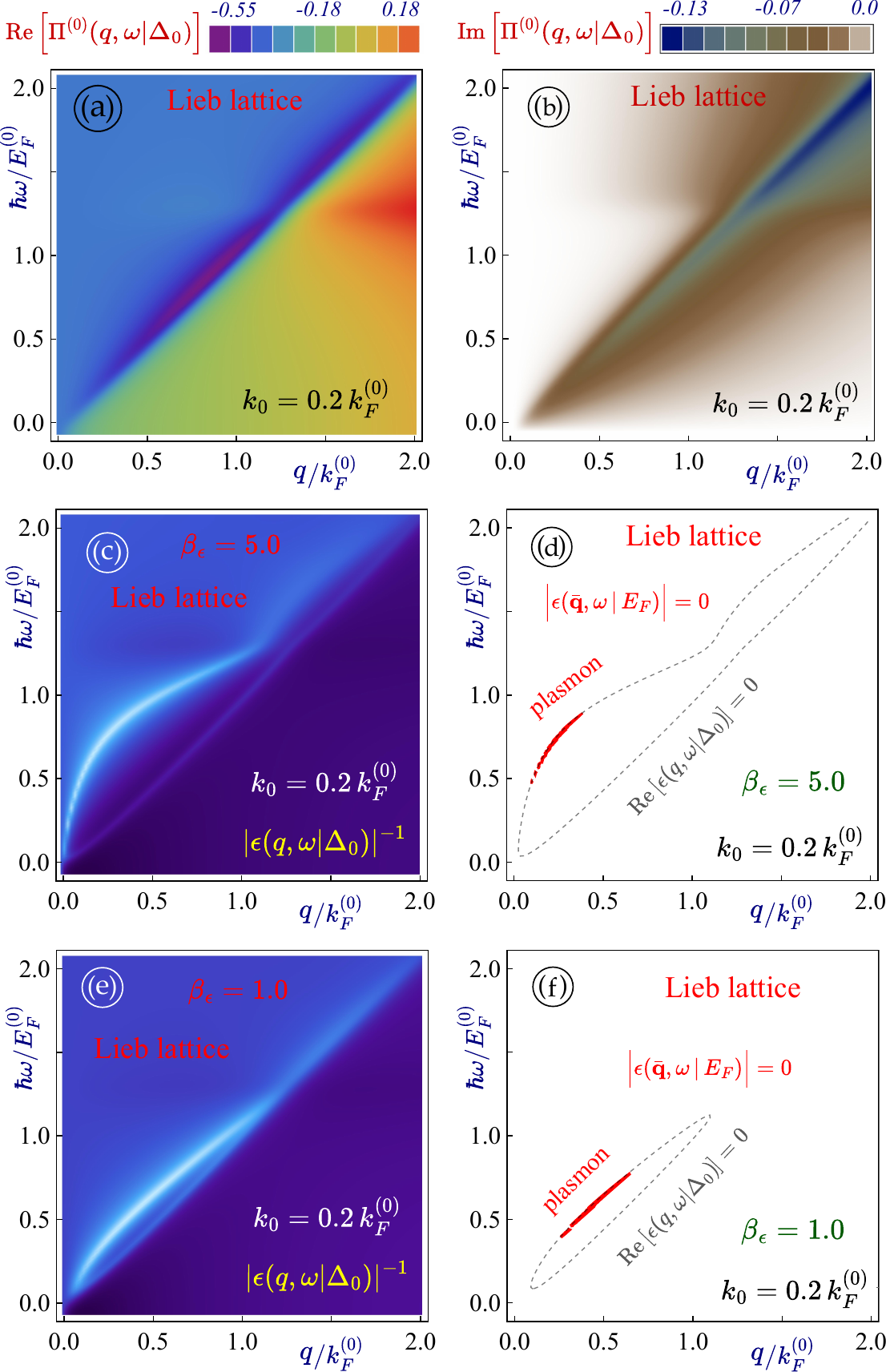}
\caption{(Color online) Polarization function and plasmon spectrum for a gapped Lieb lattice with a large gap parameter $k_0 = 0.2\,E_F^{(0)}/(\hbar v_F)$. Top panels $(a)$ and $(b)$ demonstrate the density plots of the real and imaginary parts of the polarization function $\Pi^{(0)}(q, \omega\, \vert \, k_0)$ as functions of wave vector $q$ and the plasmon excitation frequency $\omega$. Plasmon damping and singe-particle excitation regions are shown as areas with a finite imaginary part of the dynamical polarization function $\text{Im}\,[\Pi^{(0)}((q, \omega\, \vert \, k_0))]$ in panel $(b)$. The remaining plots $(c)$-$(f)$ demonstrate the plasmon spectrum in a dice lattice, obtained as the zeros of dielectric function in Eq.~\eqref{eps01}. The left panels $(c)$ and $(e)$ represent the density plots of inverse dielectric function $\left|1/\epsilon((q, \omega\, \vert \, k_0))\right|$ whose peaks correspond to the plasmon branches, while the right-hand-side plots $(d)$ and $(f)$ display numerically calculated plasmon dispersions, obtained as the zeros of  $\vert \epsilon(q, \omega\, \vert \, k_0)\vert$. The middle panels $(c)$ and $(d)$ correspond to inverse dielectric constant $\beta_\epsilon = 1/(4 \pi \epsilon_0 \epsilon_r)  = 5.0$, while the lower plots $(c)$ and $(d)$ to $\beta_\epsilon = 1$.} 
\label{FIG:4}
\end{figure}

\begin{figure} 
\centering
\includegraphics[width=0.55\textwidth]{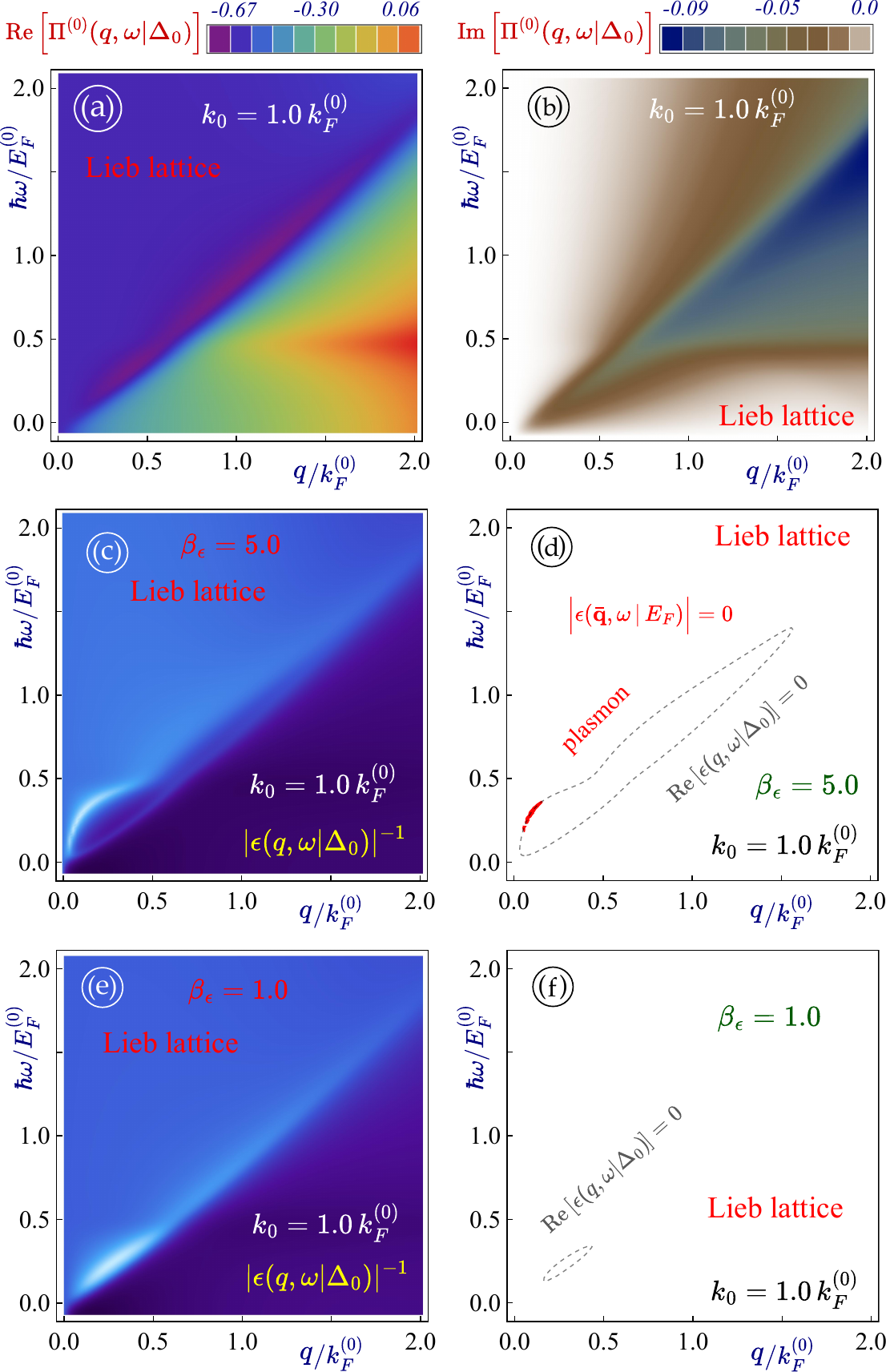}	\caption{(Color online) Polarization function and plasmon spectrum for a gapped Lieb lattice with a large gap parameter $k_0 = 0.7\,E_F^{(0)}/(\hbar v_F)$. Top panels $(a)$ and $(b)$ demonstrate the density plots of the real and imaginary parts of the polarization function $\Pi^{(0)}(q, \omega\, \vert \, k_0)$ as functions of wave vector $q$ and the plasmon excitation frequency $\omega$. Plasmon damping and singe-particle excitation regions are shown as areas with a finite imaginary part of the dynamical polarization function $\text{Im}\,[\Pi^{(0)}((q, \omega\, \vert \, k_0))]$ in panel $(b)$. The remaining plots $(c)$-$(f)$ demonstrate the plasmon spectrum in a dice lattice, obtained as the zeros of dielectric function in Eq.~\eqref{eps01}. The left panels $(c)$ and $(e)$ represent the density plots of inverse dielectric function $\left|1/\epsilon((q, \omega\, \vert \, k_0))\right|$ whose peaks correspond to the plasmon branches, while the right-hand-side plots $(d)$ and $(f)$ display numerically calculated plasmon dispersions, obtained as the zeros of  $\vert \epsilon(q, \omega\, \vert \, k_0)\vert$. The middle panels $(c)$ and $(d)$ correspond to inverse dielectric constant $\beta_\epsilon = 1/(4 \pi \epsilon_0 \epsilon_r)  = 5.0$, while the lower plots $(c)$ and $(d)$ to $\beta_\epsilon = 1$.}
\label{FIG:5}
\end{figure}

In general,  our calculated results for the energy dispersion of the plasmon modes are presented in the left/right columns of the middle/lower panels in Figs.\,\ref{FIG:2}-\ref{FIG:3} for a gapped dice lattice and in Figs.\,\ref{FIG:4}  and \ref{FIG:5} for a Lieb lattice.  The top panels in Figs.\,\ref{FIG:2} through  \ref{FIG:5}  refer to real and imaginary parts of numerically computed polarizability $\Pi^{(0)}(q, \omega\, \vert \, \Delta_0)$.  Meanwhile, we also display numerical solutions, $\omega=\Omega_{\rm pl}(q)$, of the characteristic equation in Eq.\,\eqref{eps01} for each plasmon mode so as to verify that the calculated dielectric function $\epsilon(q,\omega=\Omega_{\rm pl}(q)\,\vert\,\Delta_0)$ becomes zero within the precision of our numerical computations. For all of cases in Figs.\,\ref{FIG:2}-\ref{FIG:5}, we observe that $\Omega_{\rm pl}(q)$ for a small wave vector $q$ (long-wavelength limit) increases with $\beta_\epsilon$, implying that the plasmon energy decreases with an enhanced dielectric constant $\epsilon_r$. For a large bandgap $\Delta_0$, on the other hand, we observe an additional region for an elevated imaginary part of $\Pi^{(0)}(q, \omega\, \vert \, \Delta_0)$ above the main diagonal determined by $\omega=v_Fq$, which is further accompanied by a sign change in the real part of $\Pi^{(0)}(q, \omega\, \vert \, \Delta_0)$ and corresponds to interband transitions between the valence and conduction bands. In addition, a separate peak shows up next to the main diagonal for all accessible values of wave vector $q$ and frequency $\omega$, which is related to different interband transitions near the Fermi wave vector $k_F$. For a reduced or zero bndgap, however, both these particle-hole modes merge together and reduce to a single narrow stripe along the main diagonal $\omega = v_F q$.

\medskip

\medskip

\medskip

Specifically, our numerical results for the dynamical polarization function $\Pi^{(0)}(q, \omega\, \vert \, \Delta_0)$ and plasmon modes $\omega=\Omega_{\rm pl}(q)$ in a gapped dice lattice are presented in Figs.\,\ref{FIG:2} and \ref{FIG:3}, which correspond to a large and a small bandgap $\Delta_0$, respectively. From Fig.\,\ref{FIG:2}, we find the presence of particle-hole modes ({\em i.e.\/} regions for single-particle excitations), associated with a finite value of $\text{Im}\,[\Pi^{(0)}(q, \omega\, \vert \, \Delta_0)]$, as well as the plasmon mode slightly above the diagonal boundary with a long lifetime in regions having zero or small $\text{Im}\,[\Pi^{(0)}(q, \omega\, \vert \, \Delta_0)]$ values. Similar to a gapped graphene, for gapped dice lattices, a weakly-damped plasmon mode shows up in a larger range of wave vector $q$ below the Fermi energy $E_F^{(0)}$. Meanwhile, a particle-hole mode, which results from interband transitions from a flat band, shows up for any frequency $\hbar\omega\geq E_F^{(0)}$. Consequently, a low-damped plasmon mode could be seen only below the Fermi energy $E_F^{(0)}$. As the bandgap ratio $\Delta_0/E_F^{(0)}$ is reduced from $0.7$ to $0.2$ in Fig.\,\ref{FIG:3}, the above-diagonal bandgap-split upper particle-hole mode in Fig.\,\ref{FIG:2} is suppressed greatly.
\medskip

We present calculated polarizability and plasmon dispersion relations in Figs.\,\ref{FIG:4} -\ref{FIG:5} for a gapped Lieb lattice. Interestingly, we find that the spectra of particle-hole modes in a gapped Lieb lattice become quite different from those in a gapped dice lattice or a gapped $\alpha$-$\mc{T}_3$ model. In current case, for a small gap parameter $k_0 = \Delta_0/(2 \hbar v_F)  = 0.2\,k_F^{(0)}$ chosen for Fig.\,\ref{FIG:4}, both the electron low-energy band structure and corresponding schematics for electron transitions in a gapped Lieb lattice remain similar to those in gapless $\alpha$-$\mc{T}_3$ materials and dice lattices. Therefore, one expects to find some similarities in calculated spectra of both particle-hole and plasmon modes in these two different types of materials. 
Once the bandgap $\Delta_0$ or the parameter $k_0$ becomes large in Fig.\,\ref{FIG:5}$(b)$, one has observed quite different electronic and plasmon features in a gapped Lieb lattice. In this case, the imaginary part of polarization function exhibits a significant and extended peak in contrast to a line or narrow regions found in previously studied Dirac-cone materials. Meanwhile, we also observe an extended peak region for $\text{Re}\,[\Pi^{(0)}(q, \omega\, \vert \, \Delta_0)]$ in Fig.\,\ref{FIG:5}$(a)$. 
For dispersions of plasmon modes presented in panels $(c)$-$(f)$ of Figs.\,\ref{FIG:4} and \ref{FIG:5}, we find low-damping regions with small wave vectors $q$ for all relevant frequency ranges below the Fermi energy $E_F^{(0)}$. Therefore, we conclude that only long-wavelength plasmons could prevail with a small damping and a long lifetime. Additionally, an undamped plasmon branch could exist over a larger range of frequencies with an enhanced $\beta_\epsilon$ value since it allows higher plasmon frequencies for a fixed wave vector $q$. On the the hand, for small values of $\beta_\epsilon$  corresponding to larger dielectric constants, the plasmon modes are greatly suppressed, as seen in Fig.\,\ref{FIG:5}$(f)$.

\medskip
\par
\section{Summary and remarks}
\label{sec5}

In conclusion, we have investigated in this paper both the electronic and collective properties of pseudospin-1 Dirac materials with a flat band within their finite bandgaps. Particularly, we have calculated and compared the dynamical polarization function, particle-hole modes, as well as plasmon spectra and dampings for gapped dice and Lieb lattices, which are the most known and typical Dirac-cone materials with a flat band in their band structure and finite gaps as well among the valence, conduction and flat bands. 
In our considered models, each of these pseudospin-1 Dirac materials exhibits a very unique structure of electron transitions between empty and occupied states, which becomes adjustable by varying the Fermi energy or doping density in each material. For a dice lattice, we find a symmetric band structure in contrast to any other $\alpha$-$\mc{T}_3$ material having $0<\alpha<1$. This enables an   analytical expression for the wave-function overlap, which is a key component in calculating the polarization function. Also, we have obtained another wave-function overlaps analytically for the Lieb lattice. 

\medskip 
\par

For a dice lattice, we have found that a weakly-damped plasmon mode could exist for an extended range of wave vectors in the presence of a finite bandgap. However, this plasmon mode become strongly damped once its frequency exceeds the Fermi energy due to electron-hole mode with respect to transitions associated with a flat band. This is a genuine feature for all types of $\alpha$-$\mc{T}_3$ materials with either zero or finite bandgap. As a comparison, Lieb lattice represents a truly unique schematics for electron transitions, and its particle-hole modes reveal a large and extended peak in contrast to a narrow region found for all known Dirac-cone materials. Its plasmon modes with a long lifetime survive only for relatively small wave vectors but for a wide range of frequencies. Importantly, its plasmon modes become highly damped for all large wave vectors. As a result, it is almost impossible to observe an undamped plasmon mode in Lieb lattices with a large $k_0$ or $\Delta_0$ value.

\medskip 
\par

We are confident that the numerical results in this paper have predicted some novel electronic and collective properties of recently discovered Dirac materials, as well as new types of tunable plasmon modes and their spectra. This work will undoubtedly find numerous applications in novel nano-electronic and plasmonic devices. 

\medskip
\par

\begin{acknowledgements}
A.I. was supported by the funding received from TRADA-54-46, PSC-CUNY Award \# 66045-00 54. G.G. was supported by Grant No. FA9453-21-1-0046 from the Air Force Research Laboratory (AFRL). D.H. would like to acknowledge the Air Force Office of Scientific Research (AFOSR) and the views expressed are those of the authors and do not reflect the official guidance or position of the United States Government, the Department of Defense or of the United States Air Force.
\end{acknowledgements}
\medskip

\bibliography{DLP}
\end{document}